# Extrinsic Origin of Persistent Photoconductivity in Monolayer MoS$_2$ Field Effect Transistors

Yueh-Chun Wu[1‡], Cheng-Hua Liu[1,2‡], Shao-Yu Chen[1§], Fu-Yu Shih[1,2], Po-Hsun Ho[3], Chun-Wei Chen[3], Chi-Te Liang[2], and Wei-Hua Wang[1*]

[1]Institute of Atomic and Molecular Sciences, Academia Sinica, Taipei 106, Taiwan

[2]Department of Physics, National Taiwan University, Taipei 106, Taiwan

[3]Department of Materials Science and Engineering, National Taiwan University, Taipei 106, Taiwan

[§]Current address: Department of Physics, University of Massachusetts, Amherst, Massachusetts 01003, United States

[‡]These authors contributed equally to this work.

[*]Corresponding Author. (W.-H. Wang) Tel: +886-2-2366-8208, Fax: +886-2-2362-0200;

E-mail: wwang@sinica.edu.tw

Recent discoveries of the photoresponse of molybdenum disulfide ($MoS_2$) have shown the considerable potential of these two-dimensional transition metal dichalcogenides for optoelectronic applications. Among the various types of photoresponses of $MoS_2$, persistent photoconductivity (PPC) at different levels has been reported. However, a detailed study of the PPC effect and its mechanism in $MoS_2$ is still not available, despite the importance of this effect on the photoresponse of the material. Here, we present a systematic study of the PPC effect in monolayer $MoS_2$ and conclude that the effect can be attributed to random localized potential fluctuations in the devices. Notably, the potential fluctuations originate from extrinsic sources based on the substrate effect of the PPC. Moreover, we point out a correlation between the PPC effect in $MoS_2$ and the percolation transport behavior of $MoS_2$. We demonstrate a unique and efficient means of controlling the PPC effect in monolayer $MoS_2$, which may offer novel functionalities for $MoS_2$-based optoelectronic applications in the future.

**Introduction**

Following the discovery of graphene,[1-3] two-dimensional (2D) materials have emerged as one of the most important research topics in condensed matter physics because of the novel phenomena exhibited by and the promising applications of these materials.[4-6] In particular, semiconducting layered materials,[7,8] such as transition metal dichalcogenides (TMD),[9,10] can complement graphene because of their intrinsic bandgap and therefore can enrich the properties of these 2D materials.[11,12] Molybdenum disulfide ($MoS_2$) is a layered semiconducting TMD and therefore exhibits a bandgap,[13,14] high mobility[15,16] and strong mechanical properties.[17] Moreover, unique physical properties, including spin-valley coupling and the layer dependence of the band structure in $MoS_2$, have been demonstrated.[18-20] The combination of these interesting properties has made $MoS_2$ very attractive for new functionalities such as sensors,[21-23] logic circuits[24,25] and optoelectronic devices.[26-30]

Recently, high photoresponsivity[31,32], the photovoltaic effect[29] and the photothermoelectric effect[28] have been reported in monolayer $MoS_2$-based photodetectors and phototransistors. Optoelectronic studies on these $MoS_2$ devices demonstrated persistent photoconductivity (PPC), which is sustained conductivity after illumination is terminated.[32-35] However, a detailed understanding of PPC and its mechanism in $MoS_2$ is still not available. The origin of the trap states in $MoS_2$, which lead to the PPC effect, remains under debate.[36-39] Moreover, the PPC effect modifies the transport properties of $MoS_2$ samples, which are sensitive to the history of photon irradiation. Therefore, it is essential to understand the PPC effect to control transport phenomena in $MoS_2$.

Here, we present a systematic study of PPC in monolayer $MoS_2$ field effect transistors. The PPC dependence on the temperature, the photon dose, and the excitation energy enabled us to attribute the PPC in $MoS_2$ to random localized potential fluctuations that hinder the recombination of photoexcited electron-hole pairs. Comparing the PPC in suspended and substrate-supported $MoS_2$ devices led us to conclude that PPC originates primarily from extrinsic sources. Moreover, we could correlate PPC phenomena with percolation transport, whereby carriers transfer among nearest low-potential puddles.

Details on the fabrication of the $MoS_2$ field effect transistors can be found in the Supporting Information S1. In brief, we mechanically exfoliated $MoS_2$ flakes onto an octadecyltrichlorosilane (OTS)-functionalized $SiO_2$/Si substrate, thus obtaining a hydrophobic surface that minimized the amount of charged impurities in the 2D materials.[40] Monolayer $MoS_2$ flakes were identified by optical microscopy and subsequently confirmed using Raman spectroscopy and photoluminescence (PL) measurements (Supporting Information S1). We deposited electrical contacts (Au, 50 nm) in a two-probe geometry using a residue-free approach[41,42] to minimize contamination from the conventional lithography process. The $MoS_2$ samples were stored under vacuum for 12 hours before performing the measurements to reduce the number of adsorbed molecules (Supporting Information

S2). This procedure enabled us to investigate the photoresponse of $MoS_2$ without interference from the gas adsorbate effect.[43]

**Results and discussion**

Figure 1a shows the two-probe transconductance of sample A as a function of the back-gate voltage ($G - V_G$). The $MoS_2$ device exhibited typical n-type channel characters with a mobility of 0.7 cm[2]/Vs and an on/off ratio of $2 \times 10^4$. The $MoS_2$ transistor was then illuminated by a laser (wavelength = 532 nm) with a spot size of $\simeq 1.5$ µm. After the illumination was terminated, the conductance of the device was greatly enhanced and remained in a high-conductivity state for a time period from 2 minutes to 2 hours, depending on the irradiated photo-dose. It is noted that the longer the illumination time, the higher was the conductance.

We subsequently investigated the PPC in the $MoS_2$ transistor in greater detail. Figure 1b shows the temporal evolution of the source–drain current ($I_{DS} - t$) in vacuum (Supporting Information S3). The fast response of the current (stage 2) was attributed to a band-to-band transition that created conducting electrons and holes.[26,27] After the initial upsurge from the dark current ($I_{dark}$) level, the $I_{SD}$ increased gradually to over 2 orders of magnitude (stage 3) above the dark current. This slow increase in the $I_{SD}$ could not be attributed to the common band-to-band transition. After the laser irradiation was terminated, the $I_{SD}$ exhibited a rapid drop because of the band-to-band transition (stage 4), followed by noticeable PPC (stage 5). The photocurrent at which PPC began to dominate is denoted by $I_0$. This PPC effect was consistently observed in all of our 10 $MoS_2$ devices.

For $MoS_2$ under ambient conditions, a short photoresponse time below 1 sec has been previously reported[26] and attributed to the presence of gas adsorbates.[43] However, adsorbates were not a significant factor in the present study because our $MoS_2$ samples were treated in vacuum.

We first discuss the temperature dependence of PPC in the $MoS_2$ devices, which provides important indications of the PPC mechanism.[44] Figure 2a is a comparison of three characteristic PPC relaxations for sample A at $T = 80$, 180, and 300 K. For purposes of comparison, the dark level was subtracted, and the PPC was normalized by $I_0$. It is noted that the PPC was more pronounced at high temperatures, although the photocurrent relaxed very fast and the PPC was considerably weakened at low temperatures (Supporting Information S4). The temperature dependence of the PPC was measured by heating the samples up to room temperature, allowing the carriers to relax to equilibrium,[38,45] and then cooling the samples down in the dark to the target temperature. The PPC relaxation curves were well-fitted by a single stretched exponential decay,[46]

$$I_{PPC}(t) = I_0 \exp[-(t/\tau)^{\beta}], \tag{1}$$

where $\tau$ is the decay time constant, and $\beta$ is the exponent (Supporting Information S5). The stretched exponential decay has been widely used to model relaxation processes in complex and slowly relaxing materials.[44-47] Therefore, this photocurrent decay suggests that PPC is related to disorders in the $MoS_2$ devices.

The fitted $\tau$ and $\beta$ values at different temperatures are shown in Figure 2b. Clearly, the MoS$_2$ device exhibited a shorter $\tau$ at low temperatures, which is a signature of the random local potential fluctuations (RLPF) model.[45,48] In the RLPF model, local potential fluctuations arise either because of intrinsic disorders in the materials or extrinsic charged impurities. Therefore, low-energy electrons and holes become localized in potential minima and are spatially separated, resulting in a long recombination lifetime. The thermal excitation of carriers to higher energy states above the mobility edge produces a photocurrent after the irradiation is terminated, resulting in the PPC effect. At low temperature, the carriers are well-confined inside charge traps and therefore, negligible PPC is observed. This RLPF model has been widely used to explain PPC in II-VI compound semiconductors.[45,47,48] However, this model has not been used to explain the PPC effect in 2D TMD materials.

The PPC in the MoS$_2$ samples was attributed to the RLPF model in which the potential fluctuations originate from extrinsic[36,49] or intrinsic[50,51] sources. Extrinsic adsorbates or chemical impurities in the vicinity of the MoS$_2$ samples can lead to trap states.[36,49] Alternatively, sulfur vacancies[37] and the formation of MoO$_3$ in bulk MoS$_2$ have intrinsic causes. We determined the origin of the potential fluctuations by investigating the substrate effect on the PPC. We first measured the PPC in suspended MoS$_2$ devices that were fabricated by exfoliating monolayer MoS$_2$ onto SiO$_2$/Si substrates with trenches (width $\simeq$2 μm). Figure 3a shows the optical images of a fabricated suspended MoS$_2$ device before and after deposition of the electrodes. To verify the suspension of the MoS$_2$ sample, we measured the PL spectra of the sample, which are very sensitive to the existence of SiO$_2$/Si substrates.[52] Figure 3b is a comparison of the PL spectra of a suspended MoS$_2$ device (sample B) and a SiO$_2$-supported MoS$_2$. The PL spectrum of the suspended MoS$_2$ exhibited a strong exciton peak (A), along with trion (A$^-$) and exciton (B) peaks. The presence of the pronounced exciton peak A indicated that the MoS$_2$ sample was free from the strong n-type doping effect of the SiO$_2$ substrate,[52] indicating a suspended structure. In contrast, only trion (A$^-$) and exciton (B) peaks were observed in the PL spectrum of SiO$_2$-supported MoS$_2$.

Next, we discuss the photoresponse of the suspended MoS$_2$ (sample B) that is shown in Figure 3c. Interestingly, sample B exhibited a negligible PPC effect at $T = 300$ K, in contrast to the photoresponse of the substrate-supported MoS$_2$ at the same temperature. Only the photoresponse from the band-to-band transition was observed in this suspended MoS$_2$ device. The absence of the PPC effect in this control sample clearly indicated that the potential fluctuations in our MoS$_2$ devices had extrinsic sources. We also compared the PPC effect for MoS$_2$ devices that were fabricated on an OTS-functionalized SiO$_2$ surface and those that were fabricated on conventional SiO$_2$ substrates, as shown in Figure 3d. The PPC effect in the MoS$_2$/SiO$_2$ device was stronger than that in the MoS$_2$/OTS/SiO$_2$ device. This substrate effect was consistently observed in several samples, suggesting that there were fewer charge traps in the MoS$_2$/OTS/SiO$_2$ devices than in the MoS$_2$/SiO$_2$ device. This difference was reasonable because the OTS-functionalized SiO$_2$ surface is known to be hydrophobic and could therefore reduce surface adsorbates,[40] resulting in smaller potential variations. Further study is required to identify these extrinsic sources in detail, e.g., gas adsorbates and/or chemical impurities on the SiO$_2$ interface. Nevertheless, our finding demonstrates the importance of extrinsic sources, thereby providing a means of eliminating the PPC effect to control the photoresponse in TMD materials.

Next, we present the PPC dependence on the photon dose and the excitation energy to further validate the RLPF mechanism in our MoS$_2$ samples. Figure 4a shows the excitation power dependence of $\tau$ at room temperature, where the PPC increases with the excitation power. In the RLPF model, more carriers are excited under higher photon doses, and more electrons and holes can redistribute to occupy the sites of the local potential minima. This redistribution of carriers is therefore enhanced under high excitation power, resulting in a larger $\tau$ after the photoexcitation is terminated.[47] Figure 4b shows the PPC relaxation for different illumination times at room temperature, where $\tau$ increases with the illumination time (Supporting Information S6), Similar to the effect of excitation power, this PPC dependence on the photon dose in MoS$_2$ devices is in good agreement with the RLPF mechanism.

We further investigated the PPC of monolayer MoS$_2$ at different excitation energies ranging from 1.46 eV to 2.75 eV (Supporting Information S3), which included the optical bandgap in monolayer MoS$_2$ $\simeq 1.8$ eV.[13,14] Figure 4c is a comparison of the temporal evolution of the photoresponse for excitation energies below ($E_{ex} = 1.55$ eV) and above ($E_{ex} = 1.91$ eV) the bandgap. While the $I_{DS} - t$ curve for $E_{ex} = 1.91$ eV exhibited a typical photoresponse (which is similar to that in Figure 1b), the photoresponse was insignificant for $E_{ex} = 1.55$ eV. Figure 4d shows the excitation energy dependence of the PPC build-up level ($I_{build-up}$), which is defined as $I_0 - I_{dark}$. Here, $I_{build-up}$ represents the charging process in which photoexcited carriers fill up the local potential minimum during illumination. The photoresponse was clearly activated at $E_{ex} \sim 1.8$ eV, showing that the bandgap of monolayer MoS$_2$ is related to the PPC.[53] In the RLPF model, $I_{build-up}$ is initiated by the photoexcited carriers through the band-to-band transition, which are subsequently confined by the trap states. Therefore, PPC can only occur when the photon energy surpasses the bandgap. To briefly summarize, our observations of the PPC dependence on the temperature, the substrate effect, the photon dose, and the excitation energy fully support the RLPF model as the mechanism of PPC in the monolayer MoS$_2$ devices.

In addition to RLPF, two other mechanisms, large lattice relaxation (LLR)[54,55] and the microscopic barrier (MB)[56,57], are well-known mechanisms for PPC in a variety of materials, including mixed crystals, semiconductors, and heterostructures. In the LLR model, electrons are photoexcited from deep-level traps, and an energy barrier prevents the recapture of the electrons,[54,58] resulting in the PPC effect. Because the recapture is a thermally activated process, the PPC due to LLR is more pronounced at low temperatures, which is inconsistent with our observations (Figure 2). Another feature of the PPC that results from LLR is that a photocurrent can be activated by excitation from deep-level traps to conduction bands[59] with energies below that of the bandgap, which also contradicts the observed dependence of the PPC on the excitation energy. The MB is another mechanism for PPC in which photoexcited electron–hole pairs are spatially separated by a macroscopic potential barrier, followed by charge accumulation or trapping by barriers/spacers.[56,57] Recent studies on various structures, including $MoS_2$/graphene,[30] quantum-dot/graphene[60] and chlorophyll/graphene[61] heterostructures, demonstrated noticeable PPC due to this MB model. However, there were no such artificially created structures in our $MoS_2$ devices that could yield a macroscopic potential barrier. The PPC effect in the MB model also follows a single-exponential decay,[56] unlike the stretched exponential decay that was observed in our samples. We therefore conclude from our experimental data that LLR and MB are not mechanisms for PPC in our $MoS_2$ devices.

Finally, we consider the connection between the PPC effect and the transport properties of the $MoS_2$ devices. Figure 5a is an Arrhenius plot of the conductance of sample C at different $V_G$ values and shows that insulating behavior ( $dG/dT > 0$ ) was found over the temperature range $80 < T < 300$ K. For $T > 200$ K, the $MoS_2$ sample exhibited thermally activated behavior, where the activation energy was extracted from the linear fit ( $E_a = 98$ meV at $V_G = 0$ V). The carrier transport in this temperature regime could be described by a percolation model, in which conduction occurs via a network of spatially distributed charge puddles.[62] The activation energy then corresponded to the average potential barrier of the charge puddles. For $80$ K $< T < 200$ K, the correlation of the puddles decreased because of the decrease in the thermal energy, and the carriers could only conduct by tunneling between localized states.[62] Therefore, there was a phase transition from localized to percolation transport. The transition temperature of sample C was $T_C \sim 200$ K, which corresponded to the temperature at which the $G - 1/T$ curve deviated from thermally activated behavior (more details are provided in Supporting Information S7). To compare the temperature dependence of the PPC and transport properties, we plot $I_{build-up}$ as a function of temperature in Figure 5b. The PPC is insignificant for $T < 200$ K because of the low conductivity in the localized transport regime. Notably, $I_{build-up}$ exhibited an activated behavior for $T > 200$ K, where the transition temperature coincided with the $T_C$ value that was extracted from the transport behavior. $I_{build-up}$ could be described by a percolation approach:[44]

$$I_{build-up} \propto (T - T_C)^{\mu},$$  (2)

where $\mu$ is the characteristic exponent. We found that this function fits the data reasonably well ($\mu = 2.6$), indicating that the PPC relaxation was consistent with the percolation transport picture.

In conclusion, we have demonstrated that PPC in monolayer $MoS_2$ can be controlled by temperature, the photon dose, the excitation energy, and the substrate effect. These characteristics show that PPC in $MoS_2$ can be well-explained by the RLPF model. The potential fluctuations could be attributed to extrinsic sources because of the absence of PPC in the suspended $MoS_2$ devices. Moreover, the temperature dependence of the PPC could be described as a transition from localization to percolation models, which was in agreement with the transport properties of $MoS_2$. The results of this study can provide insight into PPC phenomena in monolayer $MoS_2$, which is important for the development of $MoS_2$-based optoelectronic applications.

**Methods**

**Sample preparation**. $MoS_2$ flakes (SPI Supplies) were mechanically exfoliated onto octadecyltrichlorosilane (OTS) self-assembled monolayer (SAM) functionalized $SiO_2$ (300 nm)/Si substrates. The surface of the OTS-functionalized $SiO_2$/Si substrate was hydrophobic with a typical contact angle above 110°. First, the $MoS_2$ flakes were identified and characterized under an optical microscope using variations in contrast and then were examined by Raman and photoluminescence spectra. We adopted resist-free fabrication to prevent contamination of the $MoS_2$ samples from the resist residue of the conventional lithography process. We used nanowire as a shadow mask to deposit metallic contacts (Au, 50 nm) with an electron-beam evaporator at a base pressure of $1.0 \times 10^{-7}$ Torr. The $MoS_2$ devices were then transferred into a cryostat (Janis Research Company, ST-500) for electrical and optical characterization. The samples were stored under a high vacuum of $1 \times 10^{-6}$ Torr to minimize the undesirable adsorption of chemical substances. A Keithley 237 was used to perform the DC electrical measurements, and a Keithley 2400 was used to apply the back gate voltage.

**References**

1       Novoselov, K. S. *et al.* Electric Field Effect in Atomically Thin Carbon Films. *Science* **306**, 666-669, doi:10.1126/science.1102896 (2004).

2       Novoselov, K. S. *et al.* Two-dimensional gas of massless Dirac fermions in graphene. *Nature* **438**, 197-200, doi:Doi 10.1038/Nature04233 (2005).

3       Zhang, Y., Tan, Y.-W., Stormer, H. L. & Kim, P. Experimental observation of the quantum Hall effect and Berry's phase in graphene. *Nature* **438**, 201-204, (2005).

4       Novoselov, K. S. *et al.* A roadmap for graphene. *Nature* **490**, 192-200 (2012).

5       Butler, S. Z. *et al.* Progress, Challenges, and Opportunities in Two-Dimensional Materials Beyond Graphene. *Acs Nano* **7**, 2898-2926, doi:Doi 10.1021/Nn400280c (2013).

6       Huang, X., Qi, X. Y., Boey, F. & Zhang, H. Graphene-based composites. *Chem. Soc. Rev.* **41**, 666-686, doi:Doi 10.1039/C1cs15078b (2012).

7       Novoselov, K. S. *et al.* Two-dimensional atomic crystals. *Proceedings of the National Academy of Sciences of the United States of America* **102**, 10451-10453, doi:DOI 10.1073/pnas.0502848102 (2005).

8       Coleman, J. N. *et al.* Two-Dimensional Nanosheets Produced by Liquid Exfoliation of Layered Materials. *Science* **331**, 568-571, doi:10.1126/science.1194975 (2011).

9       Radisavljevic, B., Radenovic, A., Brivio, J., Giacometti, V. & Kis, A. Single-layer MoS2 transistors. *Nat Nanotechnol* **6**, 147-150, doi:Doi 10.1038/Nnano.2010.279 (2011).

10      Podzorov, V., Gershenson, M. E., Kloc, C., Zeis, R. & Bucher, E. High-mobility field-effect transistors based on transition metal dichalcogenides. *Appl. Phys. Lett.* **84**, 3301-3303, doi:Doi 10.1063/1.17236395 (2004).

11      Geim, A. K. & Grigorieva, I. V. Van der Waals heterostructures. *Nature* **499**, 419-425, doi:Doi 10.1038/Nature12385 (2013).

12      Wang, Q. H., Kalantar-Zadeh, K., Kis, A., Coleman, J. N. & Strano, M. S. Electronics and optoelectronics of two-dimensional transition metal dichalcogenides. *Nat. Nanotechnol.* **7**, 699-712 (2012).

13      Mak, K. F., Lee, C., Hone, J., Shan, J. & Heinz, T. F. Atomically Thin MoS$_2$: A New Direct-Gap Semiconductor. *Phys. Rev. Lett.* **105**, 136805, doi:Artn 136805 Doi 10.1103/Physrevlett.105.136805 (2010).

14      Splendiani, A. *et al.* Emerging Photoluminescence in Monolayer MoS2. *Nano Lett.* **10**, 1271-1275, doi:Doi 10.1021/Nl903868w (2010).

15      Baugher, B. W. H., Churchill, H. O. H., Yang, Y. F. & Jarillo-Herrero, P. Intrinsic Electronic Transport Properties of High-Quality Monolayer and Bilayer MoS2. *Nano Lett.* **13**, 4212-4216, doi:Doi 10.1021/Nl401916s (2013).

16      Radisavljevic, B. & Kis, A. Mobility engineering and a metal–insulator transition in monolayer MoS2. *Nat. Mater.* **12**, 815-820, doi:10.1038/nmat3687 (2013).

17      Castellanos-Gomez, A. *et al.* Elastic Properties of Freely Suspended MoS2 Nanosheets. *Adv. Mater.* **24**, 772-775, doi:10.1002/adma.201103965 (2012).

18    Mak, K. F., He, K., Shan, J. & Heinz, T. F. Control of valley polarization in monolayer MoS2 by optical helicity. *Nat. Nanotechnol.* **7**, 494-498, (2012).

19    Xiao, D., Liu, G. B., Feng, W. X., Xu, X. D. & Yao, W. Coupled Spin and Valley Physics in Monolayers of $MoS_2$ and Other Group-VI Dichalcogenides. *Phys. Rev. Lett.* **108**, 196802, doi:Artn 196802
Doi 10.1103/Physrevlett.108.196802 (2012).

20    Zeng, H., Dai, J., Yao, W., Xiao, D. & Cui, X. Valley polarization in MoS2 monolayers by optical pumping. *Nat. Nanotechnol.* **7**, 490-493, (2012).

21    He, Q. *et al.* Fabrication of Flexible MoS2 Thin-Film Transistor Arrays for Practical Gas-Sensing Applications. *Small* **8**, 2994-2999, doi:10.1002/smll.201201224 (2012).

22    Li, H. *et al.* Fabrication of Single- and Multilayer MoS2 Film-Based Field-Effect Transistors for Sensing NO at Room Temperature. *Small* **8**, 63-67, doi:10.1002/smll.201101016 (2012).

23    Zhu, C. F. *et al.* Single-Layer MoS2-Based Nanoprobes for Homogeneous Detection of Biomolecules. *J. Am. Chem. Soc.* **135**, 5998-6001, doi:Doi 10.1021/Ja4019572 (2013).

24    Radisavljevic, B., Whitwick, M. B. & Kis, A. Integrated Circuits and Logic Operations Based on Single-Layer MoS2. *Acs Nano* **5**, 9934-9938, doi:10.1021/nn203715c (2011).

25    Wang, H. *et al.* Integrated Circuits Based on Bilayer MoS2 Transistors. *Nano Lett.* **12**, 4674-4680, doi:10.1021/nl302015v (2012).

26    Yin, Z. Y. *et al.* Single-Layer $MoS_2$ Phototransistors. *ACS Nano* **6**, 74-80, doi:Doi 10.1021/Nn2024557 (2012).

27    Lee, H. S. *et al.* $MoS_2$ Nanosheet Phototransistors with Thickness-Modulated Optical Energy Gap. *Nano Lett.* **12**, 3695-3700, doi:Doi 10.1021/Nl301485q (2012).

28    Buscema, M. *et al.* Large and Tunable Photothermoelectric Effect in Single-Layer MoS2. *Nano Lett.* **13**, 358-363, doi:10.1021/nl303321g (2013).

29    Fontana, M. *et al.* Electron-hole transport and photovoltaic effect in gated MoS2 Schottky junctions. *Sci. Rep.* **3** (2013).

30    Roy, K. *et al.* Graphene-MoS2 hybrid structures for multifunctional photoresponsive memory devices. *Nat Nanotechnol* **8**, 826-830, doi:Doi 10.1038/Nnano.2013.206 (2013).

31    Lopez-Sanchez, O., Lembke, D., Kayci, M., Radenovic, A. & Kis, A. Ultrasensitive photodetectors based on monolayer MoS2. *Nat. Nanotechnol.* **8**, 497-501, doi:10.1038/nnano.2013.100 (2013).

32    Zhang, W. *et al.* High-Gain Phototransistors Based on a CVD MoS2 Monolayer. *Adv. Mater.* **25**, 3456-3461, doi:10.1002/adma.201301244 (2013).

33    Lopez-Sanchez, O., Lembke, D., Kayci, M., Radenovic, A. & Kis, A. Ultrasensitive photodetectors based on monolayer MoS2. *Nat Nanotechnol* **8**, 497-501, doi:Doi 10.1038/Nnano.2013.100 (2013).

34    Cunningham, G. *et al.* Photoconductivity of solution-processed MoS2 films. *J Mater Chem C* **1**, 6899-6904, doi:Doi 10.1039/C3tc31402b (2013).

35    Cho, K. *et al.* Gate-bias stress-dependent photoconductive characteristics of multi-layer MoS2 field-effect transistors. *Nanotechnology* **25**, doi:Artn 155201

Doi 10.1088/0957-4484/25/15/155201 (2014).

36      Ghatak, S., Pal, A. N. & Ghosh, A. Nature of Electronic States in Atomically Thin MoS2 Field-Effect Transistors. *Acs Nano* **5**, 7707-7712, doi:Doi 10.1021/Nn202852j (2011).

37      Qiu, H. *et al.* Hopping transport through defect-induced localized states in molybdenum disulphide. *Nat. Commun.* **4**, doi:10.1038/ncomms3642 (2013).

38      Zhu, W. *et al.* Electronic transport and device prospects of monolayer molybdenum disulphide grown by chemical vapour deposition. *Nat. Commun.* **5**, doi:10.1038/ncomms4087 (2014).

39      Ghatak, S. & Ghosh, A. Observation of trap-assisted space charge limited conductivity in short channel MoS2 transistor. *Appl. Phys. Lett.* **103**, doi:Artn 122103

Doi 10.1063/1.4821185 (2013).

40      Chen, S.-Y., Ho, P.-H., Shiue, R.-J., Chen, C.-W. & Wang, W.-H. Transport/Magnetotransport of High-Performance Graphene Transistors on Organic Molecule-Functionalized Substrates. *Nano Lett.* **12**, 964-969, doi:10.1021/nl204036d (2012).

41      Cheng, H.-C., Shiue, R.-J., Tsai, C.-C., Wang, W.-H. & Chen, Y.-T. High-Quality Graphene p−n Junctions via Resist-free Fabrication and Solution-Based Noncovalent Functionalization. *Acs Nano* **5**, 2051-2059, doi:10.1021/nn103221v (2011).

42      Shih, F. Y. *et al.* Residue-free fabrication of high-performance graphene devices by patterned PMMA stencil mask. *Aip Adv* **4**, doi:Artn 067129

Doi 10.1063/1.4884305 (2014).

43      Kyungjune, C. *et al.* Gate-bias stress-dependent photoconductive characteristics of multi-layer MoS 2 field-effect transistors. *Nanotechnology* **25**, 155201 (2014).

44      Jiang, H. X. & Lin, J. Y. Percolation transition of persistent photoconductivity in II-VI mixed crystals. *Phys Rev Lett* **64**, 2547-2550, doi:DOI 10.1103/PhysRevLett.64.2547 (1990).

45      Jiang, H. X. & Lin, J. Y. Persistent Photoconductivity and Related Critical Phenomena in Zn0.3cd0.7se. *Phys Rev B* **40**, 10025-10028, doi:DOI 10.1103/PhysRevB.40.10025 (1989).

46      Palmer, R. G., Stein, D. L., Abrahams, E. & Anderson, P. W. Models of Hierarchically Constrained Dynamics for Glassy Relaxation. *Phys. Rev. Lett.* **53**, 958-961 (1984).

47      Dissanayake, A. S., Huang, S. X., Jiang, H. X. & Lin, J. Y. Charge Storage and Persistent Photoconductivity in a Cds0.5se0.5 Semiconductor Alloy. *Phys Rev B* **44**, 13343-13348, doi:DOI 10.1103/PhysRevB.44.13343 (1991).

48      Jiang, H. X. & Lin, J. Y. Percolation transition of persistent photoconductivity in II-VI mixed crystals. *Phys. Rev. Lett.* **64**, 2547-2550 (1990).

49      Lu, C. P., Li, G. H., Mao, J. H., Wang, L. M. & Andrei, E. Y. Bandgap, Mid-Gap States, and Gating Effects in MoS2. *Nano Lett.* **14**, 4628-4633, doi:Doi 10.1021/Nl501659n (2014).

50      McDonnell, S., Addou, R., Buie, C., Wallace, R. M. & Hinkle, C. L. Defect-Dominated Doping and Contact Resistance in MoS2. *Acs Nano* **8**, 2880-2888, doi:Doi 10.1021/Nn500044q (2014).

51    Ghatak, S., Mukherjee, S., Jain, M., Sarma, D. & Ghosh, A. Microscopic Origin of Charged Impurity Scattering and Flicker Noise in MoS2 field-effect Transistors. *arXiv preprint arXiv:1403.3333* (2014).

52    Scheuschner, N. *et al.* Photoluminescence of freestanding single- and few-layer MoS2. *Phys Rev B* **89**, doi:Artn 125406

Doi 10.1103/Physrevb.89.125406 (2014).

53    Wu, C.-C. *et al.* Elucidating the Photoresponse of Ultrathin MoS2 Field-Effect Transistors by Scanning Photocurrent Microscopy. *J. Phys. Chem. Lett.* **4**, 2508-2513, doi:10.1021/jz401199x (2013).

54    Lang, D. V. & Logan, R. A. Large-Lattice-Relaxation Model for Persistent Photoconductivity in Compound Semiconductors. *Phys. Rev. Lett.* **39**, 635-639, doi:DOI 10.1103/PhysRevLett.39.635 (1977).

55    Lang, D. V., Logan, R. A. & Jaros, M. Trapping Characteristics and a Donor-Complex (Dx) Model for the Persistent Photoconductivity Trapping Center in Te-Doped $Al_xGa_{1-X}As$. *Phys Rev B* **19**, 1015-1030, doi:DOI 10.1103/PhysRevB.19.1015 (1979).

56    Queisser, H. J. & Theodorou, D. E. Decay kinetics of persistent photoconductivity in semiconductors. *Phys. Rev. B* **33**, 4027-4033 (1986).

57    Theis, T. N. & Wright, S. L. Origin of Residual Persistent Photoconductivity in Selectively Doped $GaAs/Al_xGa_{1-X}As$ Heterojunctions. *Appl. Phys. Lett.* **48**, 1374-1376, doi:Doi 10.1063/1.97028 (1986).

58    Nelson, R. J. Long-Lifetime Photoconductivity Effect in N-Type Gaaias. *Appl. Phys. Lett.* **31**, 351-353, doi:Doi 10.1063/1.89696 (1977).

59    Lee, Y. C. *et al.* Observation of persistent photoconductivity in 2H-MoSe2 layered semiconductors. *J. Appl. Phys.* **99**, 063706-1-063706-4, (2006).

60    Konstantatos, G. *et al.* Hybrid graphene-quantum dot phototransistors with ultrahigh gain. *Nat. Nanotechnol.* **7**, 363-368, (2012).

61    Chen, S.-Y. *et al.* Biologically inspired graphene-chlorophyll phototransistors with high gain. *Carbon* **63**, 23-29, (2013).

62    Lo, S. T. *et al.* Transport in disordered monolayer MoS2 nanoflakes-evidence for inhomogeneous charge transport. *Nanotechnology* **25**, doi:Artn 375201

Doi 10.1088/0957-4484/25/37/375201 (2014).

**Acknowledgments**

This work was supported by Academia Sinica and the Ministry of Science and Technology of

Taiwan (Grant No. MOST 103-2112-M-001-020-MY3).

**Author Contributions Statement**

W.H.W. supervised the project. Y.C.W., C.H.L., and S.Y.C. designed the experiments. P.H.H. and C.W.C. provided the OTS-functionalized substrates. F.Y.S. provided the substrates for suspended samples. Y.C.W. and C.H.L. prepared the samples and carried out the photoresponse and transport measurements. Y.C.W., C.H.L., and S.Y.C. analyzed the data. W.H.W, Y.C.W., C.W.C., and C.T.L. wrote the paper. All authors discussed the results and contributed to the refinement of the paper.

**Additional information**

**Competing financial interests:** The authors declare no competing financial interests.

**Figure Legends**

**Figure 1. The PPC effect in a monolayer MoS$_2$ transistor.** (a) Transconductance as a function of $V_G$ in the dark and after illumination at room temperature in vacuum, showing that the conductance of the device is greatly enhanced after illumination and remains in a high-conductivity state for a long period of time. Inset: a schematic of a MoS$_2$ device on an OTS-functionalized substrate. (b) The photoresponse of the MoS$_2$ device for $V_G = 0\,\text{V}$ and $V_{DS} = 50\,\text{mV}$, which can be classified into 5 stages. In addition to the photoresponse due to band-to-band transition (stages 2 and 4), the device exhibits a slow increase in the photocurrent under illumination (stage 3) and the PPC effect (stage 5).

**Figure 2. The temperature dependence of PPC.** (a) Three characteristic PPC relaxations (circles) at $T = 80$, 180, and 300 K. PPC is more pronounced at higher temperatures but is greatly suppressed at $T = 80$ K. The PPC relaxations are well described by a stretched exponential decay (solid line). The stretched exponential decay is $I_{PPC}(t) = I_0 \exp[-(t/\tau)^\beta]$ with the dark current subtracted. $I_0$ in the PPC relaxation curves is normalized to 1 for comparison of the decay rate. (b) The temperature dependence of the decay time constant ($\tau$) and the exponent ($\beta$). The grey region indicates the temperature range over which PPC is not observed.

**Figure 3. The substrate effect of PPC.** (a) Upper panel: optical images of the suspended MoS$_2$ device before and after deposition of the electrode. Lower panel: a schematic of the suspended monolayer MoS$_2$ device. (b) PL spectra of suspended (upper panel) and SiO$_2$-supported (lower panel)

monolayer MoS$_2$ devices. (c) Photoresponse of the suspended MoS$_2$ device showing negligible PPC at $T = 300$ K ( $V_G = 0$ V, $V_{DS} = 5$ V). (d) Photoresponses of monolayer MoS$_2$ on OTS-functionalized and conventional SiO$_2$ substrates ($V_G = 0$ V, $V_{DS} = 5$ V). The photocurrents are normalized by $I_0$ for purposes of comparison.

**Figure 4. Photon dose and excitation wavelength dependences of the PPC relaxation.** (a) Excitation power dependence of the decay time constant for $V_G = -60$ V and $V_{DS} = 50$ mV. (b) The PPC relaxations for different illumination times (10 seconds and 120 seconds) at $V_G = \pm 60$ V and $V_{DS} = 50$ mV. (c) The temporal evolution of $I_{SD}$ for excitation energies below ( $E_{ex} = 1.55$ eV) and above ( $E_{ex} = 1.91$ eV) the bandgap at $T = 300$ K and $V_G = 0$ V. (d) The PPC build-up level ( $I_{build-up}$ ) at different excited photon energies and $V_G = 0$ V. The photon dose and excitation energy dependences of the PPC are consistent with the RLPF model in the monolayer MoS$_2$ devices.

**Figure 5. Correlations between PPC and transport behaviors.** (a) The Arrhenius plot for the conductance of sample C at different $V_G$ values. The solid lines indicate linear fits in the thermally activated regime. (b) The temperature dependence of the PPC buildup level ( $I_{build-up}$ ) at $V_G = 0$ V. For $T > 200$ K, $I_{build-up}$ is well described by a percolation model, which is shown as a solid line. The shaded region denotes the temperature regime over which transport is dominated by tunneling between localized states.

(a)

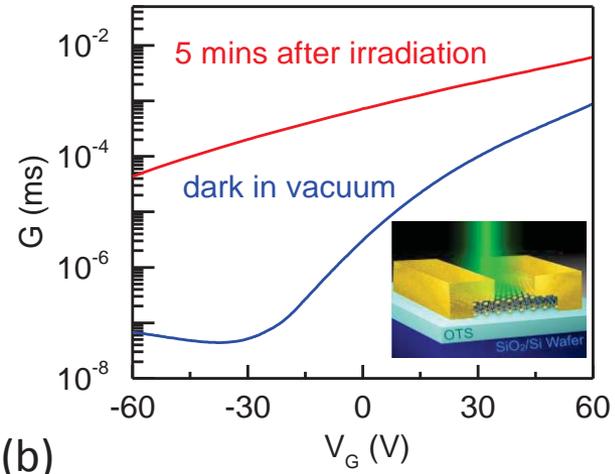

(b)

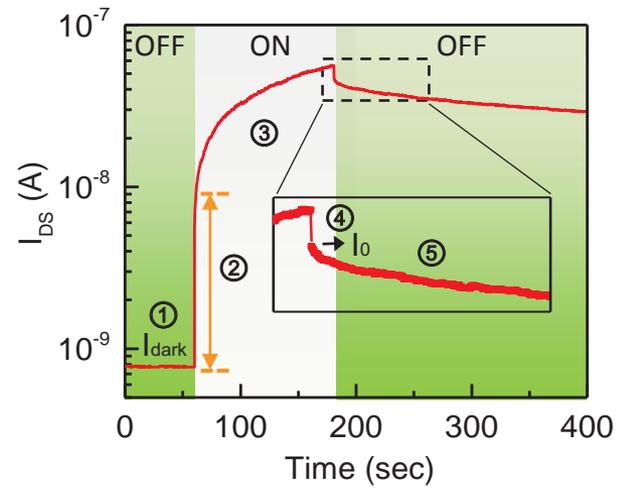

(a)

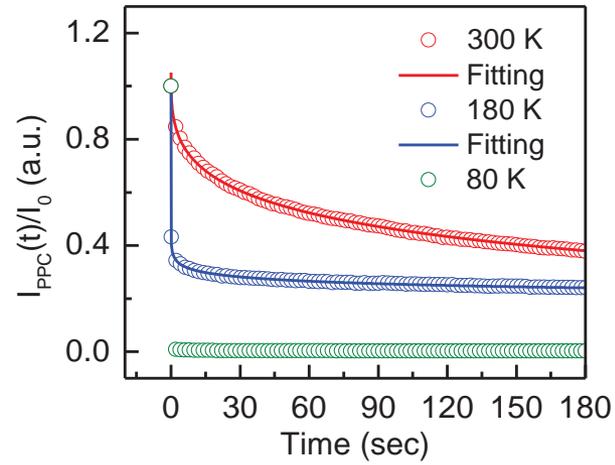

(b)

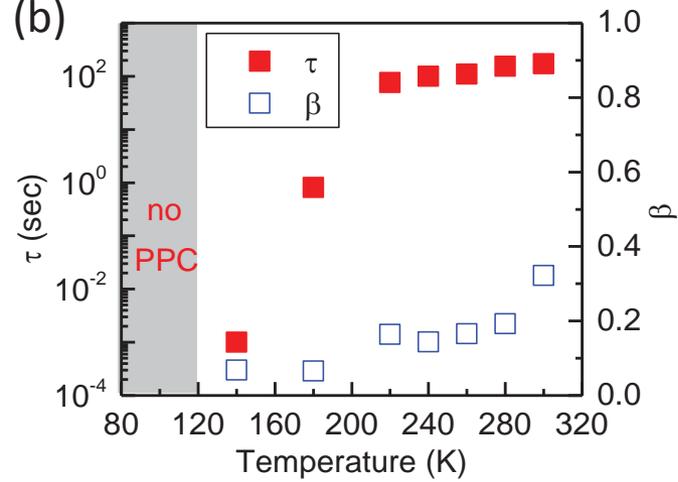

(a)

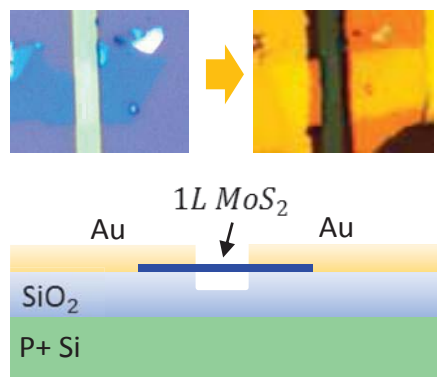

(b)

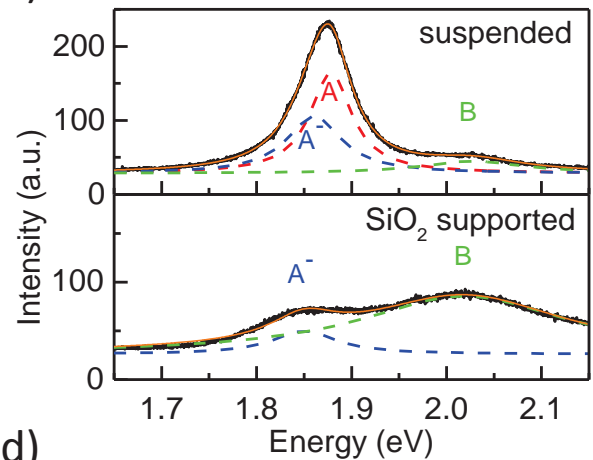

(c)

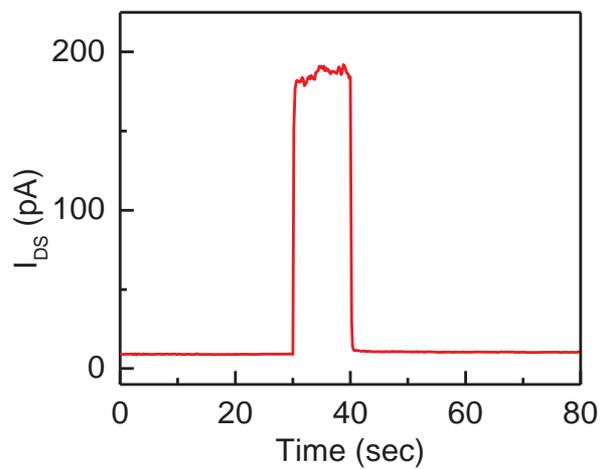

(d)

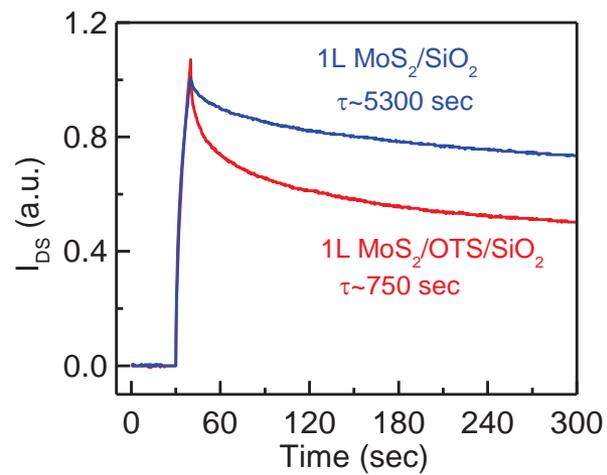

(a)
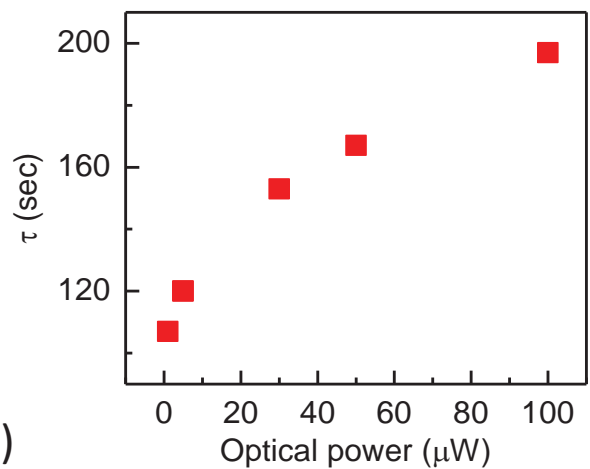

(b)
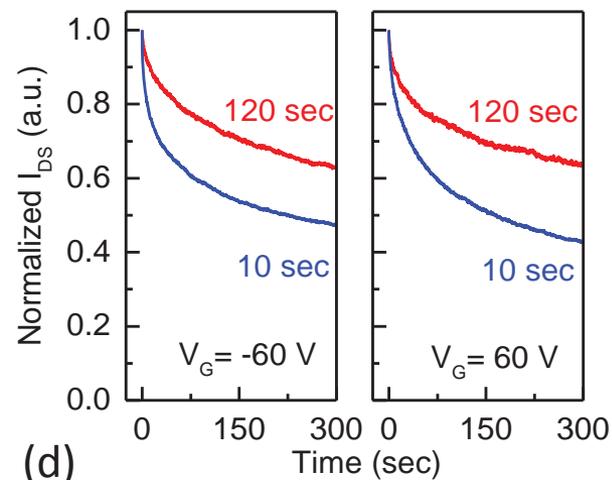

(c)
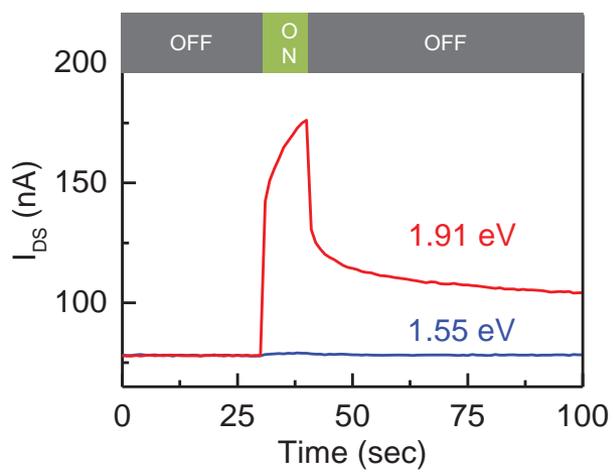

(d)
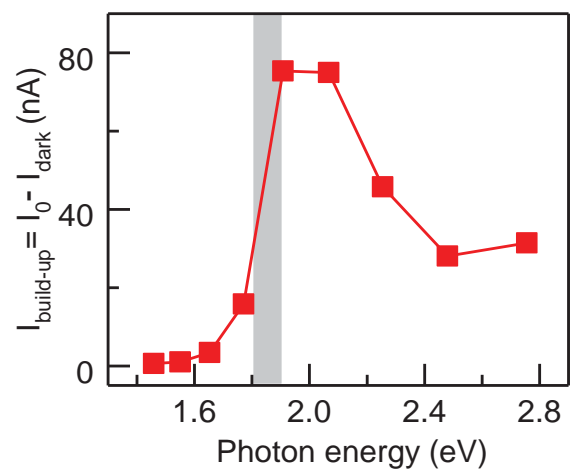

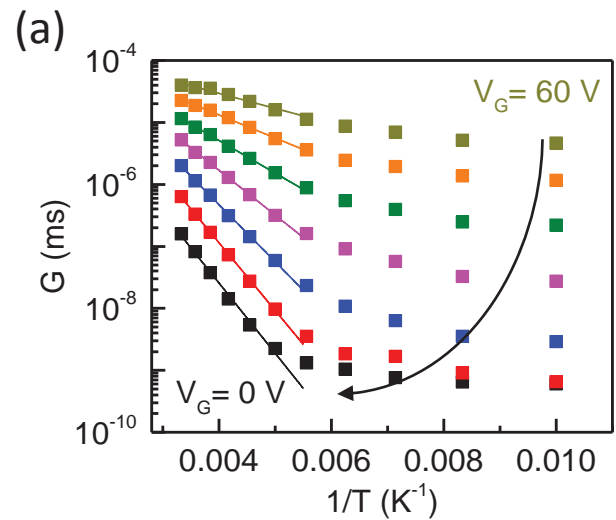

(a)

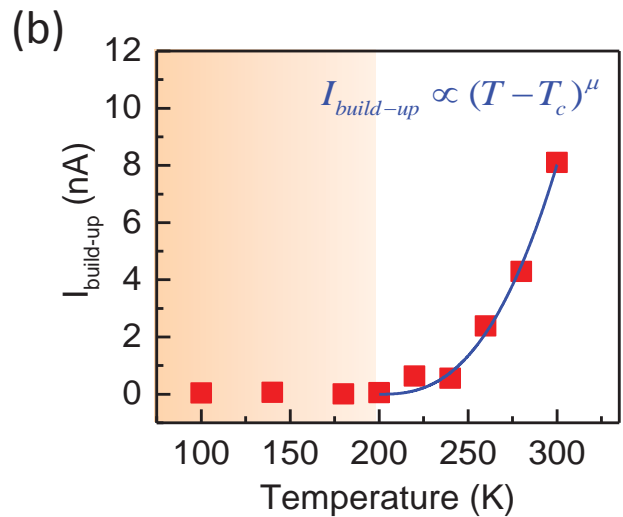

(b)

$I_{build-up} \propto (T - T_c)^\mu$

# Extrinsic Origin of Persistent Photoconductivity in Monolayer MoS₂ Field Effect Transistors

Yueh-Chun Wu[1‡], Cheng-Hua Liu[1,2‡], Shao-Yu Chen[1§], Fu-Yu Shih[1,2], Po-Hsun Ho[3], Chun-Wei Chen[3], Chi-Te Liang[2], and Wei-Hua Wang[1*]

[1]Institute of Atomic and Molecular Sciences, Academia Sinica, Taipei 106, Taiwan

[2]Department of Physics, National Taiwan University, Taipei 106, Taiwan

[3]Department of Materials Science and Engineering, National Taiwan University, Taipei 106, Taiwan

[§]Current address: Department of Physics, University of Massachusetts, Amherst, Massachusetts 01003, United States

[‡]These authors contributed equally to this work.

[*]Corresponding Author. (W.-H. Wang) Tel: +886-2-2366-8208, Fax: +886-2-2362-0200;

## S1. Device Fabrication

MoS$_2$ flakes (SPI Supplies) were mechanically exfoliated onto octadecyltrichlorosilane (OTS) self-assembled monolayer (SAM) functionalized SiO$_2$ (300 nm)/Si substrates. The surface of the OTS-functionalized SiO$_2$/Si substrate was hydrophobic with a typical contact angle above 110°. The hydrophobic surface decreased the number of absorbate molecules, thereby decreasing the charged-impurity scattering in MoS$_2$ and charge traps.[1] First, the MoS$_2$ flakes were identified and characterized under an optical microscope using variations in contrast. Figure S1a is a typical optical microscopy image of the MoS$_2$ sample after deposition of the electrode. Figure S1b is the photoluminescence (PL) spectrum of a typical monolayer and bilayer MoS$_2$ sample that were fabricated on OTS-functionalized substrates, showing the exciton peaks A and B. Due to the difference in the quantum efficiency, the intensity of the exciton peaks A and B in the monolayer MoS$_2$ are larger than those in the bilayer MoS$_2$ samples. Another PL peak (I) at $\simeq 1.6$ eV corresponding to the indirect interband transition was observed in bilayer MoS$_2$. Figure S1c shows a Raman spectrum (blue curve) of a monolayer MoS$_2$ sample with two characteristic peaks at 388.7 cm$^{-1}$ and 407.0 cm$^{-1}$ that correspond to the E$_{2g}$ and A$_{1g}$ resonance modes. The difference between the two peaks is $\simeq 18.3$ cm$^{-1}$, which is consistent with that obtained for the monolayer MoS$_2$ from previous reports. For comparison, the Raman spectra of the bilayer and trilayer MoS$_2$ samples are also shown.

We adopted resist-free fabrication to prevent contamination of the MoS$_2$ samples from the resist residue of the conventional lithography process. We used nanowire as a shadow mask to deposit metallic contacts (Au, 50 nm) with an electron-beam evaporator at a base pressure of $\simeq 1.0 \times 10^{-7}$ Torr. The MoS$_2$ channel length was $\simeq 1$ μm. The MoS$_2$ devices were then transferred into a cryostat (Janis Research Company, ST-500) for electrical and optical characterization. The samples were stored under a high vacuum of $1 \times 10^{-6}$ Torr to minimize the undesirable adsorption of chemical substances. A Keithley 237 was used to perform the DC electrical measurements, and a

Keithley 2400 was used to apply the back gate voltage. Figure S1d shows the source-drain current ($I_{DS}$) as a function of the source-drain bias ($V_{DS}$), which is linear over the small bias region.

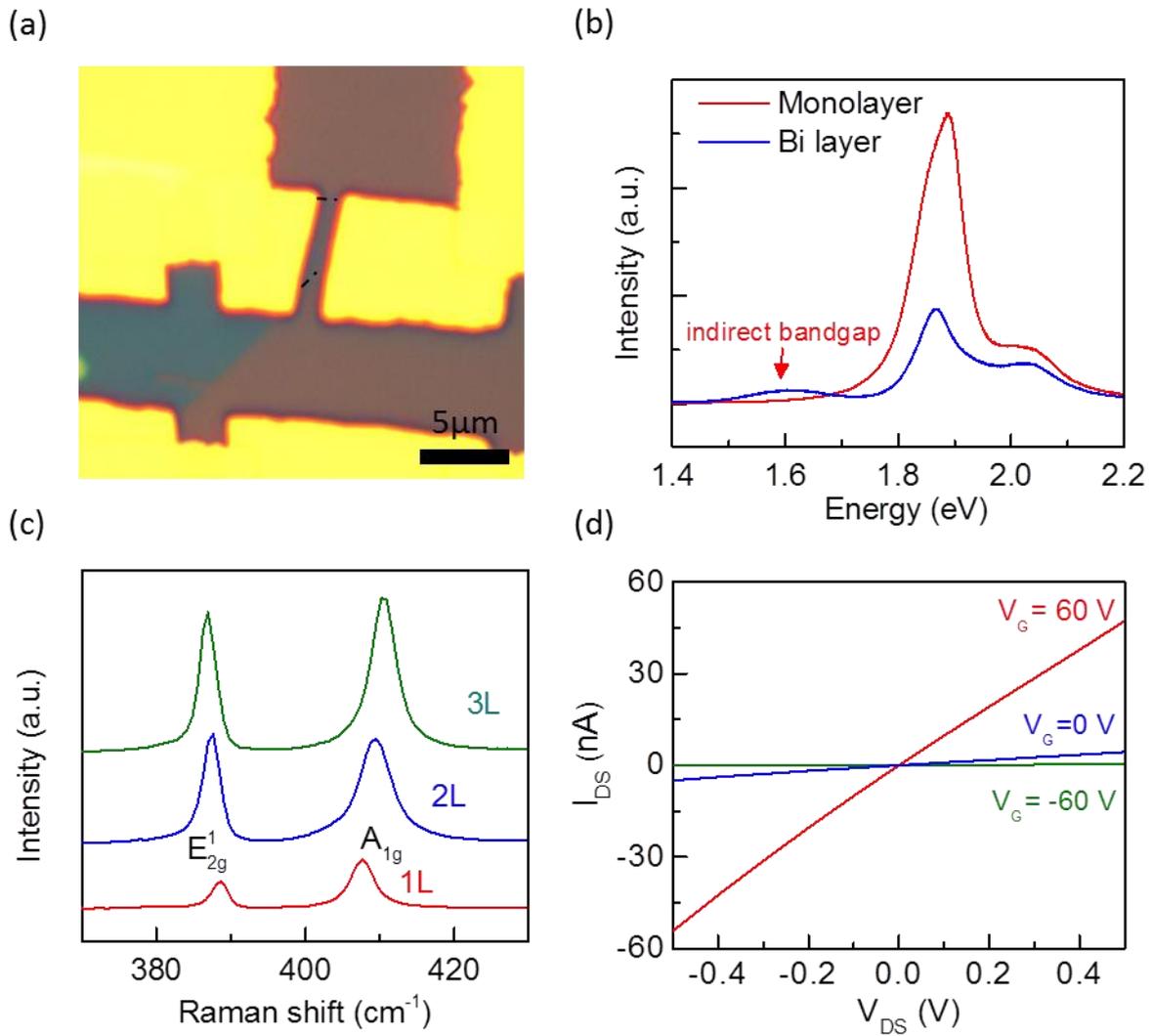

**Figure S1.** (a) Optical image of MoS$_2$/OTS FET devices (sample A) discussed in the main text. The channel length is $\simeq 1$ μm. (b) PL spectra of monolayer and bilayer MoS$_2$ on OTS-functionalized substrates. (c) Raman spectra of monolayer, bilayer, and trilayer MoS$_2$ on OTS-functionalized substrates, showing E$_{2g}$ and A$_{1g}$ peaks of different MoS$_2$ layers. (d) The $I_{DS} - V_{DS}$ curve of the MoS$_2$/OTS/SiO$_2$ device.

## S2. Gas Adsorbate Effect

We stored the $MoS_2$ samples in vacuum ($1 \times 10^{-6}$ Torr) for 12 hours before conducting the measurements to minimize the gas adsorbate effect. Figure S2 shows the transconductance curves of the $MoS_2$ samples as a function of the back-gate voltage ($G - V_G$ curves). We investigated the effect of the gas adsorbates by comparing the $G - V_G$ curves of the as-fabricated sample under ambient conditions and of the sample that was stored in a vacuum. Under ambient conditions, the $MoS_2$ device exhibited typical n-type channel characteristics with a mobility of 0.03 $cm^2 / V \cdot s$ and an on/off ratio of $\simeq 500$. There was no significant difference between the $G - V_G$ curves before and after irradiation although a small PPC effect appeared (decay time < 10 sec). However, after illumination in a vacuum, the device showed strong persistency in the high-conductivity state for several minutes to hours. Moreover, after storing the sample in a vacuum for 12 hours, the dark current stabilized and was reversible after several cycles of illumination and PPC relaxation. The conductance increased significantly beyond its original level under ambient conditions. Previous studies have shown that the degradation in the electric performance of monolayer $MoS_2$, which was measured under ambient conditions, can be attributed to adsorbed water and oxygen molecules that act as extrinsic scattering centers.[1-4] By illumination, the gas adsorbates can be removed and pumped out in vacuum, thereby enhancing the electrical characteristics of $MoS_2$ devices.[5]

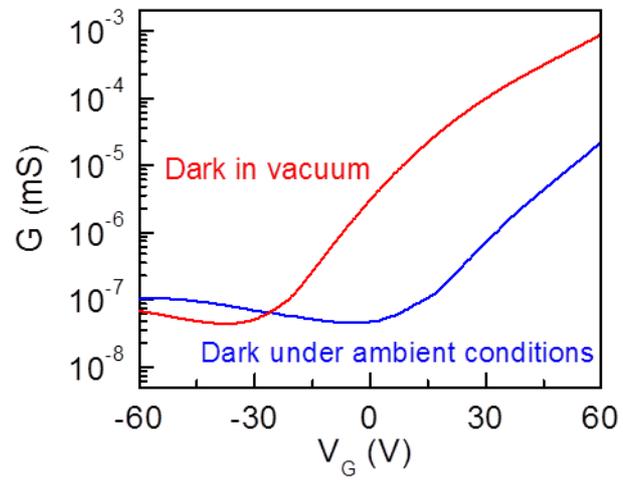

**Figure S2.** Transconductance of the monolayer $MoS_2$/OTS/$SiO_2$ device under ambient and vacuum conditions. Following illumination in vacuum, the enhanced mobility, the on/off ratio, and the threshold voltage are observed.

# S3. Photoresponse measurements

Figure S3 shows a schematic of the measurement system that integrated optical microscopy, Raman/PL spectroscopy, and the photoresponse measurements. A solid-state CW laser (Nd:YAG, 532 nm) was employed as the light source in the photoresponse measurements. The incident light was guided into the microscope and focused by an objective (10X, NA 0.3) with a spot size of $\simeq 1.5$ μm. The power density on the sample was estimated at $\simeq 5.6 \times 10^4$ W/cm$^2$ for an illumination power of 1 mW. For Raman and PL spectroscopy, a 100X objective (NA 0.6) was used with a spot size of $\simeq 0.7$ μm. A mercury-xenon lamp (Hamamatsu Photonics, 150 W) was used as the broadband white light source for the excitation energy dependence of photoresponse (see Figures 3c and 3d in the main text). The white light was selected using a band-pass filter for peak wavelengths from 450 nm to 850 nm (spacing: 50 nm; FWHM: 40 nm). The dependence of the photoresponse on the excitation energy was measured by fixing the power density at 15.9 $W/cm^2$ for different excitation energies.

We characterized the PPC relaxation by choosing the starting point of the PPC ($I_0$ in Figure 1b of main text) as the first data point after the photocurrent sharply dropped. Here, we assumed that the band-to-band recombination time was comparable to the lifetime of the PL ($\simeq 100$ ps).[6,7] Because the time delay of the PC measurement was 200 ms, which was much larger than the photocurrent relaxation due to band-to-band recombination, the photocurrent after $I_0$ was dominated by PPC.

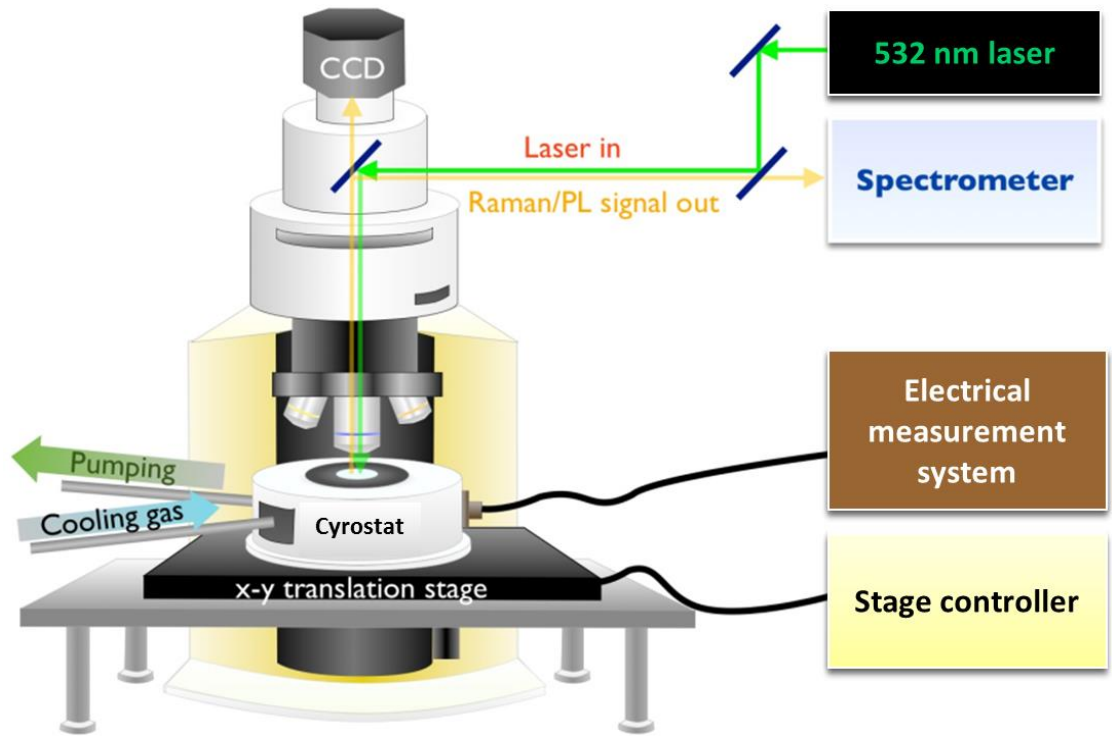

**Figure S3.** Schematic of the cryostat system with integrated OM, PL, Raman, and photocurrent

measurements.

## S4. Temperature dependence of PPC relaxation

Before each measurement, the MoS$_2$ samples were warmed up to room temperature to ensure that the dark current had reached equilibrium and then cooled down in the dark to the desired temperature. Detailed data on the temperature dependence of PPC are shown here, in which the PPC is normalized by $I_0$ for purposes of comparison. The PPC clearly weakened as the temperature decreased for samples A and B. As shown in Figure 2b in the main text, both $\tau$ and $\beta$ decreased as the temperature decreased. At low temperatures, the PPC dropped more quickly for t < $\tau$ but became more persistent for t > $\tau$, which could be accurately described by a stretched exponential function with a small $\beta$.[8]

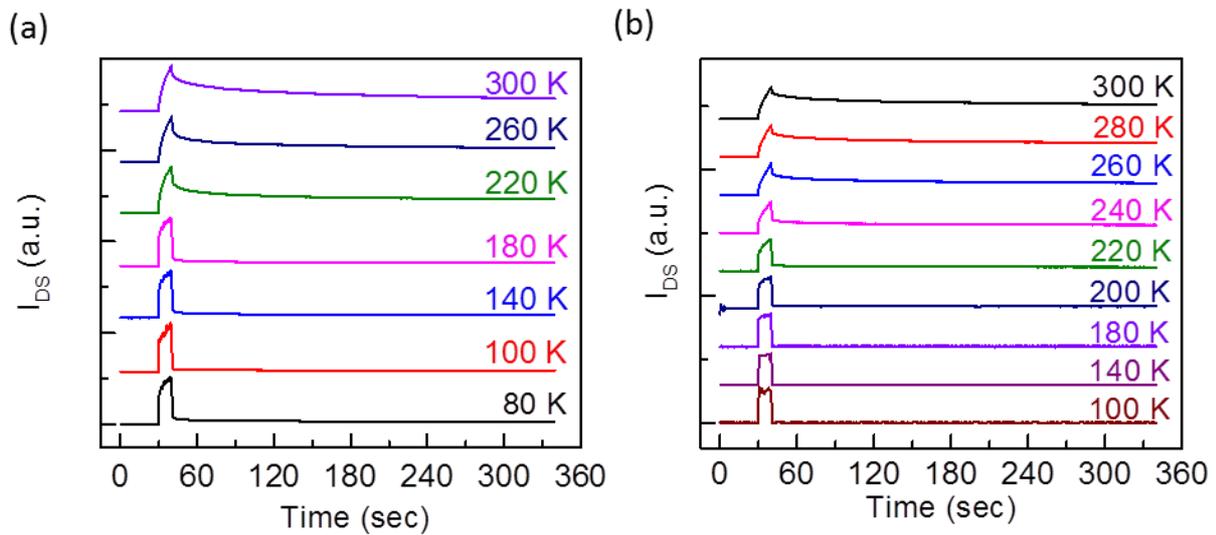

**Figure S4.** Temporal evolution of the photoresponse for (a) sample A and (b) sample B at different temperatures. The PPC is normalized by $I_0$ and offset vertically for purposes of comparison.

## S5. Fitting of the PPC relaxation curves

To verify the validity of the stretched exponential decay for describing the observed PPC in monolayer $MoS_2$, we consider other possible relaxation approaches, including single exponential, double exponential, and logarithmic decay. The fitting results of two representative PPC relaxation curves at T = 300 K and 180 K by these different schemes are shown in Figure S5. It can be seen that only the stretched exponential decay yields satisfactory fitting result for the whole temporal range at different temperatures. As mentioned in the main text, the observed stretched exponential decay of the PPC relaxation in our $MoS_2$ devices suggests a disordered system, which is consistent with the percolation transport model discussed in Figure 5 of the main text.

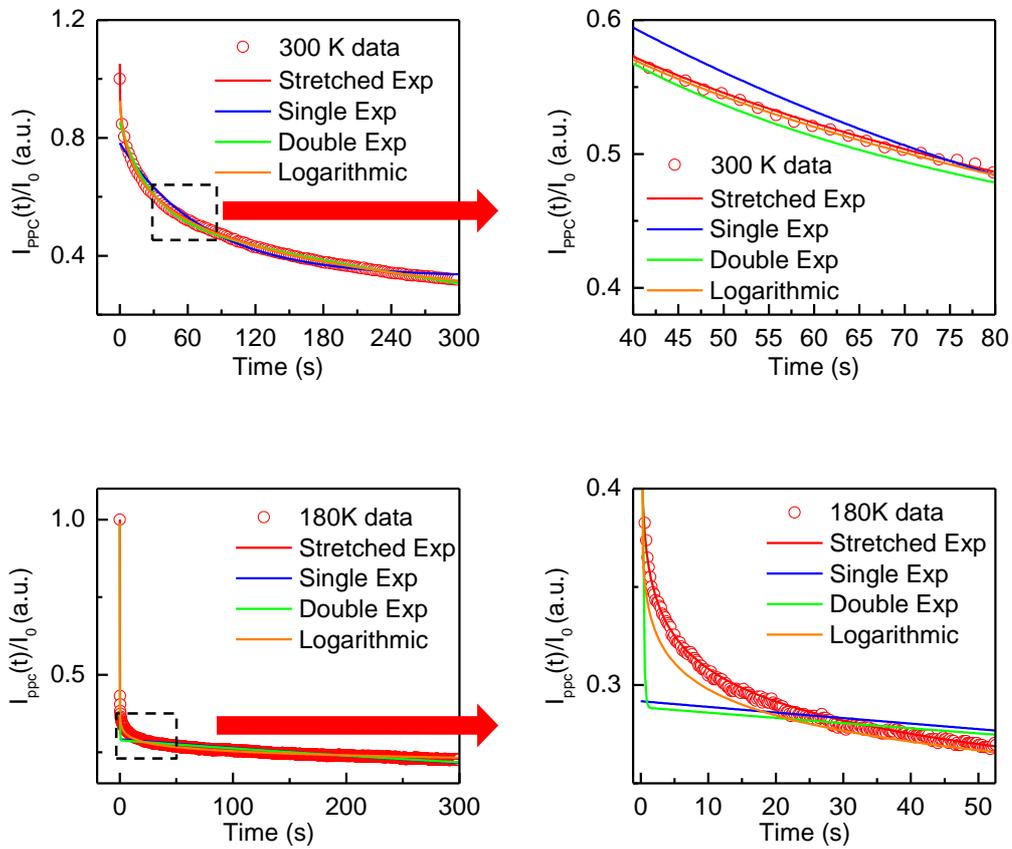

**Figure S5.** Two representative PPC relaxation curves at T = 300 K and 180 K and the fitting results with various schemes, including stretched exponential, single exponential, double exponential and logarithmic decay.

## S6. Gate voltage dependence of the PPC

We discuss the back gate voltage ($V_G$) dependence of the PPC relaxation here. We show the PPC relaxation for $V_G = -60$ V and $V_G = +60$ V in Figure S6a (also in Figure 4b of the main text), as well as the $V_G$ dependence of $\tau$ in Figure S6b. The position of Fermi level, which is controlled by $V_G$, can greatly affect the carrier density and consequently the conductance in the MoS$_2$ channel. This $V_G$ dependence of conductance is revealed in Figure 1a of the main text and the related discussion. Nevertheless, the PPC relaxation is mainly determined by the density of the trap states and the extent of the carrier trapping in the RLPF model. Because there are various possible sources of trap states, including electron traps, hole traps, and mid-gap states in MoS$_2$ samples,[9] it is plausible that the strength of the PPC is correlated to the density of the trap states as Fermi level is tuned with $V_G$. However, we note that the correlation may be complicated and further study is required to elucidate detailed $V_G$ dependence of the PPC effect.

(a)                                        (b)

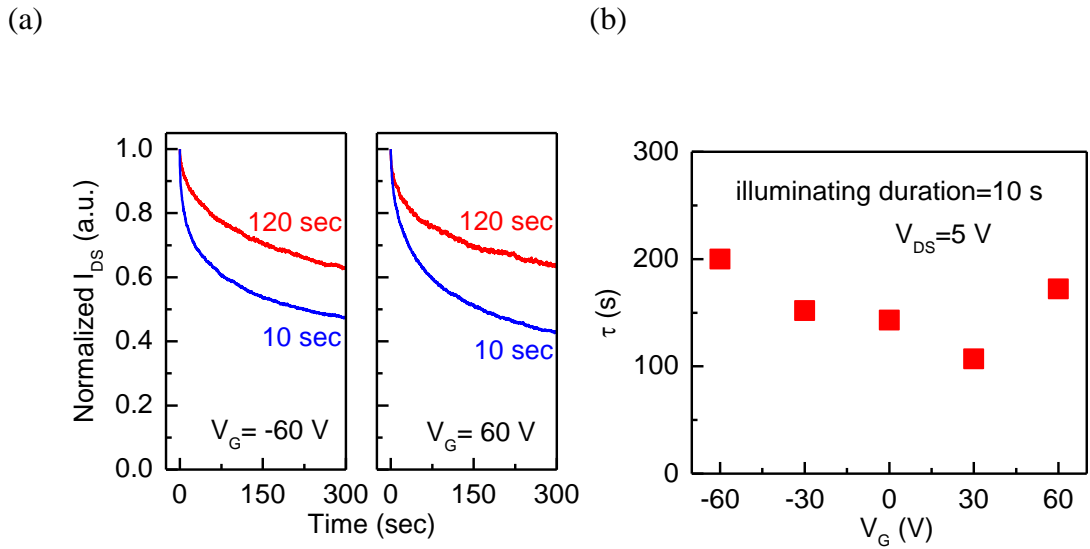

**Figure S6.** (a) The normalized PPC relaxation curves at $V_{DS} = 50$ mV under $V_G = -60$ V (left panel) and $V_G = 60$ V (right panel). Two different irradiation duration (10 sec and 120 sec) were used to excite the MoS$_2$ sample. No significant difference between the PPC relaxation for $V_G = 60$ V and $V_G = -60$ V was observed. (b) The fitted $\tau$ versus applied $V_G$ at $V_{DS} = 5$ V for illumination duration of 10 sec.

## S7. Transport characteristics of the MoS₂ FET

In addition to characterizing the transport behavior of sample B, as discussed in the main text, we show the transfer curves for sample A at different temperatures in Figure S7a. Figure S7b is Arrhenius plot for the conductance of sample A at different $V_G$ values. The transport behavior of sample A could be classified into three different regimes. (1) For $T > 240$ K, the MoS₂ sample exhibited metallic behavior where $\sigma$ decreased rapidly with the temperature. Over this temperature range, carriers above the mobility edge dominated the transport, resulting in metallic behavior. (2) For $160 < T < 240$ K, the sample exhibited thermally activated behavior. The carrier transport in this temperature regime could be described by a percolating picture in which conduction occurred via a network of spatially distributed puddles. (3) For $80 < T < 160$ K, the correlation of the puddles decreased because of the decrease in the thermal energy, and the carriers could only conduct by tunneling between localized states. Similar to sample B in the main text, the build-up level of the PPC was closely related to the carrier transport, which exhibited percolation behavior ($\mu = 2.2$) above the transition temperature $T_C = 160$ K, as shown in Figure S7c.

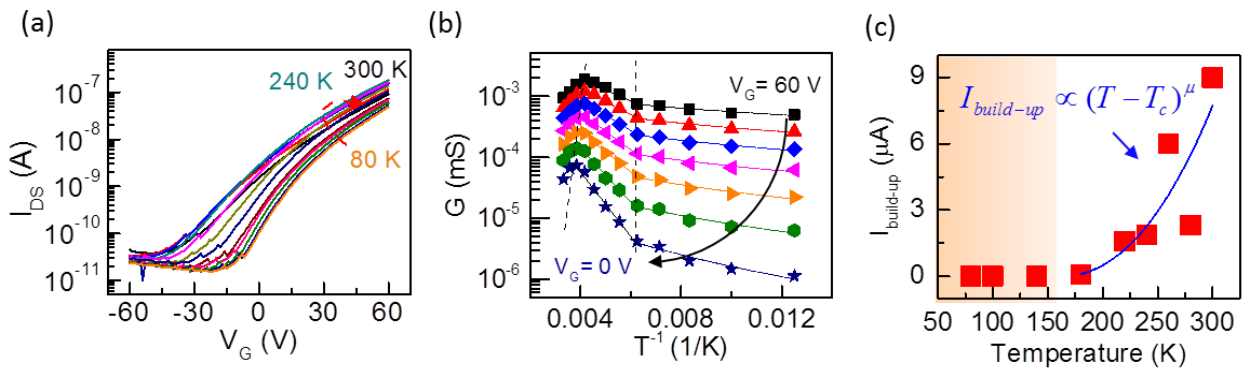

**Figure S7.** Transport characteristics and PPC for sample A: (a) The two-probe transfer curves at different temperatures. (b) Arrhenius plots of the conductance at different $V_G$ values. (c) The temperature dependence of the PPC buildup level ($I_{build-up}$), where the solid line is the fitted result using the transition temperature $T_C = 160$ K, which is obtained from the transport measurement.

**S8. Carrier mobility dependence of the PPC**

Figure S8 shows a distribution of τ versus mobility for 7 monolayer MoS$_2$ samples that were obtained by same illumination condition. The data suggests that the PPC becomes more persistent for higher mobility samples. It is conceivable that more persistent PC is caused by higher density of the trap states in the MoS$_2$ channel. This could result in shorter average distance between the localized states and greater hopping rate among these states, leading to higher mobility. However, the detail mechanism behind the relation between τ and mobility requires further study.

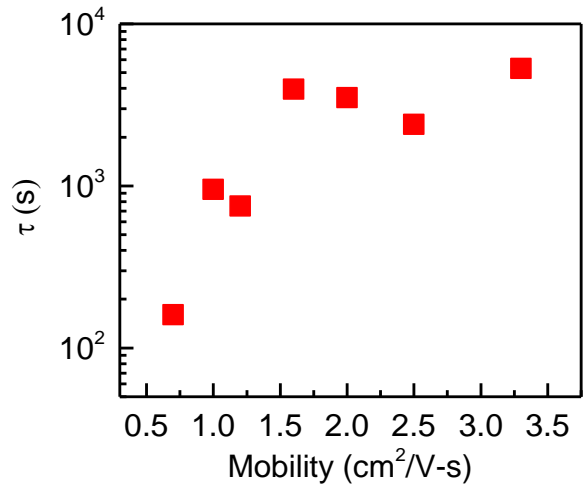

**Figure S8.** The distribution of τ versus mobility for 7 monolayer MoS$_2$ samples that were obtained by same illumination condition.

# References

1       Li, S. L. *et al.* Thickness-Dependent Interfacial Coulomb Scattering in Atomically Thin Field-Effect Transistors. *Nano Lett.* 13, 3546-3552, doi:Doi 10.1021/Nl4010783 (2013).

2       Tongay, S. *et al.* Broad-Range Modulation of Light Emission in Two-Dimensional Semiconductors by Molecular Physisorption Gating. *Nano Lett.* 13, 2831-2836, doi:10.1021/nl4011172 (2013).

3       Cho, K. *et al.* Electric Stress-Induced Threshold Voltage Instability of Multilayer MoS2 Field Effect Transistors. *Acs Nano* 7, 7751-7758, doi:10.1021/nn402348r (2013).

4       Qiu, H. *et al.* Electrical characterization of back-gated bi-layer MoS2 field-effect transistors and the effect of ambient on their performances. *Appl. Phys. Lett.* 100, 123104, doi:Artn 123104

Doi 10.1063/1.3696045 (2012).

5       Zhang, W. *et al.* High-Gain Phototransistors Based on a CVD MoS2 Monolayer. *Adv. Mater.* 25, 3456-3461, doi:10.1002/adma.201301244 (2013).

6       Korn, T., Heydrich, S., Hirmer, M., Schmutzler, J. & Schuller, C. Low-temperature photocarrier dynamics in monolayer MoS2. *Appl. Phys. Lett.* 99, 102109, doi:Artn 102109

Doi 10.1063/1.3636402 (2011).

7       Kozawa, D. *et al.* Photocarrier relaxation pathway in two-dimensional semiconducting transition metal dichalcogenides. *Nat Commun* 5, 4543, doi:10.1038/ncomms5543 (2014).

8       Johnston, D. C. Stretched exponential relaxation arising from a continuous sum of exponential decays. *Phys. Rev. B* 74, 184430 (2006).

9       Furchi, M. M., Polyushkin, D. K., Pospischil, A. & Mueller, T. Mechanisms of Photoconductivity in Atomically Thin MoS2. *Nano Lett.* 14, 6165-6170, doi:10.1021/nl502339q (2014).